# Efficient Photon Upconverters with Ionic Liquids

By *Yoichi Murakami**

[*] Assist. Prof. Y. Murakami
Global Edge Institute, Tokyo Institute of Technology
2-12-1 Ookayama, Meguro-ku, Tokyo 152-8550, Japan
E-mail: murakami.y.af@m.titech.ac.jp

## Abstract

This paper presents the development and characterization of photon upconverters fabricated with ionic liquids (ILs), which are novel fluids that are recently drawing attention due to their unique properties, such as negligible vapor pressures and high thermal stabilities. The upconverters in this study are based on triplet-triplet annihilation (TTA) between excited polycyclic aromatic molecules, and TTA requires a fluidic media to allow for the molecules to collide with each other for energy transfer. This process, TTA-based photon upconversion (TTA-UC), was therefore mainly accomplished with organic solvents previously. It is found that the molecules used for TTA-UC, which are non-polar or weakly polar, are stably solvated in a certain class of ILs. The mechanism of the observed solvation is proposed and discussed. The upconversion quantum yields (UC-QYs) measured by continuous wave (CW) light excitation reach as high as 10 % with moderate excitation intensity (~ 6 W/cm$^2$), which is considerably higher than those in previous TTA-UC studies performed with organic solvents (up to ~ 4 % with CW light excitations). It is found that the value of UC-QY starts to saturate as the excitation power increases for all the cases, even within the moderate CW excitation power range in this study. An analytical model that describes the UC-QY vs. excitation power relationship is derived and compared with the experimental results. The agreement between them suggests that the donor-acceptor energy transfer in this system is highly efficient. Based on these experimental and analytical findings, it is found that efficient energy transfer between the molecules is possible in ILs and therefore ILs are not actually viscous media for the purpose of TTA-UC. Further, the use of ILs are advantageous for TTA-UC because of the extreme ease of the removal of oxygen molecules, which efficiently quench excited triplet states, with the use of ultra-high vacuum turbo-molecular pumps exploiting the negligible vapor pressures of ILs.

## 1. Introduction

Efficient utilization of solar energy for production of secondary energies has been an important and urgent subject to the science and engineering communities. Production of secondary energy such as electrical power (by photovoltaics) and hydrogen (by photocatalytic water-splitting catalysts) utilize only a portion of the solar spectrum for energy conversion, which limits the overall efficiencies. Photons with energies below a threshold energy, which is system dependent, are not utilized in secondary power generation. Therefore, developing a method that can make use of these photons allows for higher efficiencies for light conversion systems.

To resolve this problem, one proposed strategy has been the use of photon upconversion (UC),[1] which is a process of converting two or more photons of lower energy into one photon of higher energy. Thus far, UC by rare-earth elements such as erbium and ytterbium using their successive absorption of photons has been widely studied for more than past 40 years.[2] However, this technology is usually implemented with several orders of magnitude higher light intensities ($10^1$ – $10^4$ W/cm$^2$)[2d-2i] than that of the sunlight (~ 0.1 W/cm$^2$), and UC based on rare-earth elements are currently considered mainly for biological imaging and probing applications.[2f-2i]

Recently, UC that can be implemented with light intensities as low as the solar intensity have been proposed by exploiting triplet-triplet annihilation (TTA) between excited aromatic molecules.[3] For example, Baluschev and co-workers[3a-3c] have demonstrated that UC is possible even with light intensities close to that of terrestrial solar irradiance. They combined palladium porphyrins and anthracene derivative molecules, where the former were the triplet sensitizers and the latter were the triplet reservoirs, dissolved in toluene. Since this approach employs a TTA process, the media must be fluidic in order for the molecules to diffuse and collide with each other to transfer their triplet energies efficiently. By this reason, the triplet-triplet annihilation based photon upconversion (TTA-UC) has so far been performed with organic solvents such as toluene,[3b-3d] benzene,[3e] and trichloroethane.[4] However, the use of these solvents are a significant hurdle for the realization of TTA-UC for practical applications because they are highly volatile, flammable, and often incompatible with common plastic materials.

This work presents a new class of TTA-based photon upconverters, fabricated with ionic liquids (ILs) as the fluidic host materials. Ionic liquids are novel molten salts that are in the liquid state even at room temperatures, which have recently drawn attention in materials science and technology,[5] partly because of their negligible vapor pressures[5,6] and high thermal stabilities up to several hundred degrees Celsius.[5,7] As imagined by their *ionic* nature, ILs are known to have polarities similar to those of short-chain alcohols,[8] while the polycyclic aromatic molecules used in TTA-UC are non-polar or weakly polar. This leads to an intuitive prediction that ILs are an unsuitable media for the purpose of TTA-UC. In this paper, it is demonstrated that the IL based TTA photon upconverters are not only possible and temporally stable, but also exhibit unprecedentedly high upconversion quantum yields (UC-QYs). While the values of UC-QYs thus far reported with volatile solvents are up to ~ 4 % with continuous-wave (CW) light excitations[3a-3d,9], the present work has achieved as high as 10 % UC-QY under moderate CW excitation powers (~ 6 W/cm$^2$).

The findings obtained through this study are two-fold: (i) ILs, although often regarded as viscous media, are not actually a viscous media in terms of the rate of energy transfers between molecules, and (ii) the advantage of using ILs is the compatibility with a direct evacuation of the samples by ultra-high vacuum turbo-molecular pumps for deoxygenation, compared with freeze-pump-thaw cycles or inert gas bubbling needed for volatile organic solvents.

## 2. Results and Discussion

### 2.1. Samples

The ILs tested in this report are listed in **Table 1**. Here, abbreviations of [$C_n$mim] (1-*alkyl*-3-methylimidazolium), [$C_n$dmim] (1-*alkyl*-2,3-dimethylimidazolium), and [$NC_1C_2C_2(C_2OCH_3)$] (*N*,*N*-diethyl-*N*-methyl-*N*-(2-methoxyethyl) ammonium) for cations and [$NTf_2$] (bis(trifluoromethylsulfonyl)imide) and [$CTf_3$] (tris(trifluoromethylsulfonyl)methide) for anions are used. The ILs were purchased from IoLiTec (#1 – #7, #10), Covalent Associates (#1, #2, #6, #9), Kanto Chemical (#8, #13), Merck (#11), and TCI (#12).[10] To implement TTA-UC, *meso*-Tetraphenyl-tetrabenzoporphine Palladium ($PdPh_4TBP$) and perylene (**1** and **2** in **Figure 1a**, from Sigma-Aldrich) were used as the triplet sensitizer and photon emitter, respectively. The fundamental mechanism of the TTA-UC with sensitizing and emitting molecules has been discussed previously[3a,4] and will be described in the analysis section of this paper.

It has been known that ILs, including those employed here, have polarities similar to those of short chain alcohols.[8] On the other hand, polycyclic aromatic molecules used for TTA-UC, including **1** and **2,** are non-polar or weakly polar and generally do not dissolve in methanol. **Figure 1b** is a photo taken 24 h after the powders of **1** and **2** were sprinkled over IL #1 held in a quartz mortar. After 24 h the powders were still floating on the IL's surface, while only a part of the IL around the powders was faintly colored, indicating that they hardly dissolve in the IL spontaneously.

The fabrication procedure developed for this study is described as follows. First, stock solutions of **1** and **2** in toluene (concentrations: $4 \times 10^{-4}$ M and $4 \times 10^{-3}$ M, respectively) were prepared and stored under nitrogen until just before use. The stock solutions were added with a mechanical pipette to an IL held in a glass vial, which resulted in a layer-separation (panel (i) of **Figure 2a**). Typically, the sensitizer stock (10 - 50 μl) and the emitter stock (200 – 300 μl) were added to an IL (400 μl). The ILs with high hydrophobicity (#1 – #9) have been found to be miscible with toluene in finite amounts (e.g., up to ~ 240 μl toluene miscible with 400 μl of IL #1[11]). For this class of ILs, the solutions were readily made uniform looking using shear mixing done by gentle repeated suction-and-ejection with a glass Pasture pipette (panel (ii) of Figure 2a). Immediately after this, the vial was capped and underwent ultrasonication for 10 – 20 minutes. Subsequently, the vial was opened and placed in a vacuum chamber and pumped by an oil-free scroll pump for 4 – 10 h to remove the toluene (panel (iii) of Figure 2a).[12] The vial was then set in a purpose-made high-vacuum chamber that was inside of a stainless steel (SUS) vacuum glovebox. The high-vacuum chamber was evacuated by a turbo-molecular pump for at least 12 h until it reached ultra-high vacuum range ($10^{-4}$ – $10^{-5}$ Pa), while the vapor pressure of ILs at room temperature are several orders of magnitude lower ($10^{-10}$ – $10^{-9}$ Pa[6]). Finally, the high-vacuum chamber was opened in the Ar-filled SUS glovebox, and the sample liquid was injected and sealed in an appropriate glass container depending on the purpose. For UC-QY measurements, the sample liquid was injected into a square cross-section quartz tube and sealed with solder inside the Ar-filled glovebox. Details of UC-QY measurements are described in the Experimental Section. As for the rest of the ILs shown in Table 1 (#10 – #13), which are moderately-to-highly miscible with water[5b], they have been found to be unable to form uniform mixtures with the stock solutions, and hence are not further investigated in this paper.

## 2.2. Results

**Figure 2b** shows a photo of an upconversion of CW He-Ne laser light (632.8 nm, 10 mW). Bright blue emission was clearly seen under room light illuminations. **Figure 2c** shows the emission spectrum of the same sample, ranging for 450 – 620 nm. As was shown by Figure 1b, the powders of **1** and **2** (which prefer non-polar media) show poor spontaneous dissolution into the ILs (which are moderately polar media). This causes concerns about the stability of these molecules in ILs after the removal of toluene by vacuum pumping. To examine this, the following aging experiment was performed. First, several 1 mm-thick quartz cuvettes were filled with the samples in the Ar-filled glovebox and plugged with the teflon caps. The UV-vis optical absorption spectra of these cuvettes were measured, and then again in the Ar-filled glovebox they were sealed in a purpose-made aluminum chamber. The chamber was then transferred to a forced convection oven and maintained at 80 °C for 100 h, followed by an additional 39 h held at room temperature. As soon as the lid was opened, the UV-vis absorption spectra of the cuvettes were measured. **Figures 3a – 3d** compare the optical absorption spectra before and after the aging process for the samples made with ILs #1, #2, #6, and #8, respectively. The spectra showed no change by this process, indicating that the molecules had been stably solvated in the ILs. Additional evidence for the temporal stability is shown by **Figure 3e**, where a sample made with IL #1 was left in an atmospheric environment under room light for 6 months. This sample still showed bright blue emission upon CW laser irradiation (632.8 nm, 10 mW).

The UC-QYs measured for different samples are shown in **Table 2**.[13] The excitation light source was a 632.8 nm CW laser (30 mW with a spot diameter of 0.8 mm, corresponding to ~

6 W/cm$^2$). Along the series of [$C_n$mim][$NTf_2$] (#1 – #4 for $n$ = 2, 4, 6, and 8), the UC-QY takes a maximum at $n$ = 6. Especially among ILs #1 – #3, the result is noteworthy because the ILs with longer alkyl chain, which have higher viscosities,[14] resulted in higher UC-QYs. This is different from previous findings for TTA-UC, where an increase in the media's viscosity led to a lowering of UC-QYs.[15] Additionally it is remarkable that the UC-QY is so strongly dependent on the IL. Especially for the group of [$C_n$dmim][$NTf_2$] (#5 – #7) the dependence is significant, whereas the reason has not been elucidated in this study (while reproducible[13]). Sample #7 showed the highest UC-QY of 10.6%, which is much higher than previously reported results for samples fabricated with less viscous volatile organic solvents where the UC-QYs were up to ~ 4% under CW light excitation.[3a-3d,9] **Table 3** shows the influence of the sensitizer and emitter concentrations on the UC-QY; higher emitter concentration and lower sensitizer concentration resulted in higher QY-UC. The latter may be explained in terms of a quenching of the excited emitter molecules by the ground state sensitizer molecules, as previously proposed.[16]

### 2.3. Discussion

As has been shown by Figure 3, both the sensitizer and emitter molecules **1** and **2** are stably solvated in the ILs for a long time. This opposes expectation because ILs are polar in general[8] while the aromatic molecules prefer non-polar environments. This simplified view that is sorely rooted in the consideration of solvent's polarity, however, neglects cation-π interaction[17]. I propose cation-π interaction is the mechanism for the observed stabilization of the molecules in the ILs. The emitter and sensitizer molecules are abundant in π electrons. The cation-π interaction has been proposed to be an important stabilization mechanism based on an electrostatic attraction between a positive charge and a π face of an aromatic system.[17] To examine its role in the present system, the miscibility of ILs with benzene ($C_6H_6$) and cyclohexane ($C_6H_{12}$) were compared. **Figure S1** in the Supporting Information shows that benzene was moderately miscible but cyclohexane, which is similar but lacking π electrons, is completely immiscible with the same ILs. The result implies that *an existence of π electrons in the molecules, not the extent of polarity, dominates the solvation* in the present study.

Table 2 shows that longer alkyl chain length in the cation led to higher UC-QY (for #1 – #3). The higher viscosities due to the longer alkyl chains in the cations[14] are expected to slow the diffusive motion of the molecules. Since TTA-UC relies on molecular collisions it is unexpected that the UC-QY increases with increasing alkyl chain length. To investigate further, the dependence of photoemission on the excitation power was investigated. **Figure 4a** shows the dependence of measured UC emission intensity on the excitation power, plotted on a double-logarithmic scale. At lower powers, the slopes of the plots are close to (but less than) two, and as the power increases, the slopes monotonically decrease and approach unity. Accordingly, the UC-QYs start to saturate toward their respective constant values (**Figure 4b**).

In previous TTA-UC studies measured with CW excitation (where volatile organic solvents were employed for media), the emission intensities were reported to change quadrically with the excitation powers.[3e,3f] This quadratic dependence was explained that the TTA-UC process is a two-body annihilation process and hence the chance for a triplet molecule to find an annihilation partner within its life time is proportional to the square of the created triplet concentration. Under such circumstances, the UC-QY increases linearly with the excitation power. Recently, Cheng et al.[16] have reported a quadric-to-linear transition in the photoemission intensity by exciting their toluene-based samples with intense femtosecond laser pulses (the power for the pulse duration was as high as 13 GW/cm$^2$). The authors have suggested two possibilities for the observed quadric-to-linear transition: (i) photo-bleaching caused in the sensitizing molecules by the intense pulse excitations and (ii) achievement of sufficiently high triplet densities so that virtually all of the created triplets could find their annihilation partners within their lifetimes and hence the emission intensities linearly correlated with the excitation powers. The authors stated that both mechanisms were considered to coexist in their results.[16]

In the present study, the excitation was weak CW (up to 6 W/cm$^2$) and hence the possibility of photo-bleaching has been excluded.[18] The results in Figure 4 are explained as follows, based on the second possibility suggested by Cheng et al.[16]. When the excitation power is low and hence the triplet population is low, the TTA rate is determined by the probability of triplet species to find annihilation partners. This is a competing process with their decay to the ground state, which may be caused by spontaneous intersystem crossing or by the residual oxygen molecules that efficiently quench the triplet states. However, when the excitation intensity is high enough to create sufficient spatial density of the triplets *and when the oxygen concentrations in the media is sufficiently low*, virtually all of the created triplets find annihilation partners before they decay into the ground state even with weak excitations, leading to the saturation of the UC-QY values.

As stated above, the excitation in the present study is CW and is much weaker than the previous study. In addition, the measured viscosities of the ILs are reported to be about two-orders of magnitude higher than the common volatile solvents. In the previous TTA-UC study,[16] however, the authors also mentioned the possibility of the incomplete removal of oxygen molecules from the samples. One distinct difference of the present study from the previous studies is the use of IL-based samples that can be thoroughly degassed by pumping with ultra-high vacuum turbo-molecular pumps exploiting negligible vapor pressure of ILs (typically $10^{-10}$ – $10^{-9}$ Pa). Whether or not efficient triplet energy transfer between the molecules is possible in ILs is considered based on an analytical model in the next section. However, this is already supported by the two experimental observations previously discussed, (i) that the slopes of the emission intensities approach unity as the excitation power increases as shown by Figure 4a, and (ii) that the magnitudes of the UC-QYs for [C$_n$mim][NTf$_2$] ($n$ = 2, 4, 6) do not correlate with the IL's viscosity as shown by Table 2.

### 3. Modeling and Analysis

The energy level diagram in the present study is shown by **Figure 5a**. For typical conditions, the emitter concentration is much higher than that of the sensitizer, i.e., [E] >> [S]. In palladium porphyrins the $^1S^* \rightarrow {}^3S^*$ intersystem crossing (ISC) is known to occur with almost unity quantum yield.[19] Under this circumstance and with steady CW excitation, the number conservation of the energy-carrying quanta is described for the sensitizer by the following equation, in terms of the densities of the sensitizer molecules in the lowest triplet state [$^3S^*$], the emitter molecules in the lowest triplet state [$^3E^*$], and the emitter molecules in the ground singlet state [$^1E^G$], all of which are in unit of molar (M)

$$N_{\mathrm{ex}} = k_{\mathrm{T(S)}}[^3\mathrm{S}^*] + k_{\mathrm{TET}}[^3\mathrm{S}^*][^1\mathrm{E}^{\mathrm{G}}] + k_{\mathrm{TTA}}[^3\mathrm{S}^*][^3\mathrm{E}^*]. \quad (1)$$

In Eq. (1), $N_{ex}$ is the molar rate of photons absorbed by the sensitizer (M s$^{-1}$), $k_{T(S)}$ is the triplet decay rate of the sensitizer (s$^{-1}$), $k_{TET}$ is the triplet energy transfer rate from sensitizer to emitter (M$^{-1}$ s$^{-1}$), and $k_{TTA}$ is the TTA rate between molecules in triplet states (M$^{-1}$ s$^{-1}$), as indicated in Figure 5a. On the other hand, the number conservation equation for the emitter is

$$k_{\mathrm{TET}}[^3\mathrm{S}^*][^1\mathrm{E}^{\mathrm{G}}] = k_{\mathrm{T(E)}}[^3\mathrm{E}^*] + k_{\mathrm{TTA}}[^3\mathrm{S}^*][^3\mathrm{E}^*] + 2k_{\mathrm{TTA}}[^3\mathrm{E}^*]^2, \quad (2)$$

where $k_{T(E)}$ denotes the triplet decay rate of the emitter (s$^{-1}$). From Eqs. (1) and (2), the equation

$$N_{\mathrm{ex}} = k_{\mathrm{T(E)}}[^3\mathrm{E}^*]\left(1 + \frac{k_{\mathrm{T(S)}}[^3\mathrm{S}^*]}{k_{\mathrm{T(E)}}[^3\mathrm{E}^*]}\right) + 2k_{\mathrm{TTA}}[^3\mathrm{E}^*]^2\left(1 + \frac{[^3\mathrm{S}^*]}{[^3\mathrm{E}^*]}\right) \quad (3)$$

is obtained. The rate of photon emission from the lowest exited singlet level of the emitter, denoted by $N_{em}$ (M s$^{-1}$), is expressed by the following equation

$$N_{\mathrm{em}} = \varepsilon\varphi k_{\mathrm{TTA}}[^3\mathrm{E}^*]^2, \quad (4)$$

where $\varepsilon$ and $\varphi$ denote the photoemission quantum efficiency from the $^1E^*$ level and the statistical branching ratio of the $^3E^* + {}^3E^* \rightarrow {}^1E^*$ in the TTA process, respectively ($0 \leq \varepsilon \leq 1$; $0 \leq \varphi \leq 1$). Further, the UC-QY ($\theta$: $0 \leq \theta \leq 1$) is defined as

$$\theta \equiv \frac{2N_{\text{em}}}{N_{\text{ex}}} = \frac{2\varepsilon\varphi k_{\text{TTA}}[^3\text{E}^*]^2}{N_{\text{ex}}}, \qquad (5)$$

where Eq. (4) was used to derive the right-most term.

Before proceeding further, an assumption of

$$k_{\text{TET}}[^1\text{E}^{\text{G}}] >> k_{\text{T(S)}}, \qquad (6)$$

which is the condition for a highly efficient donor-acceptor energy transfer,[20] is considered. This assumes that the energies of $^3S^*$ are transferred to the emitter more rapidly than they decay into $^1S^G$. To see if this holds, the order of $k_{TET}$, which is considered to be similar to the diffusion controlled rate constant, $k_{dif}$, due to its spin-conserving exchange mechanism (Dexter mechanism), is estimated using the following Debye Equation: [20]

$$k_{\text{TET}} \approx k_{\text{dif}} = \frac{8RT}{3000\eta}. \qquad (7)$$

In Eq. (7), $R$: the gas constant ($8.31 \times 10^7$ erg mol$^{-1}$), $T$: temperature (K), and $\eta$: viscosity of the media expressed in unit of poise (P). Since the typical viscosities of the ILs in this study fall in the range of 0.4 – 1 P,[14] the range of $k_{dif}$ calculated by Eq. (7) is $6.6 \times 10^7 - 1.7 \times 10^8$ M$^{-1}$ s$^{-1}$, and here $1 \times 10^8$ M$^{-1}$ s$^{-1}$ is taken as the representative value. Accordingly, since the typical emitter concentration in this study is $3 \times 10^{-3}$ M, $k_{TET}$ [$^1E^G$] is estimated to be ~ $3 \times 10^5$ s$^{-1}$. On the other hand, the triplet lifetime of $PdPh_4TBP$ (**1**) in polar media has been reported to be ~ 260 μs,[21] which corresponds to $k_{T(S)}$ ~ $4 \times 10^3$ s$^{-1}$. Although the $k_{T(S)}$ in ILs may not be the same rigorously speaking, as an order-of-magnitude estimation, $k_{TET}$ [$^1E^G$] is derived to be two orders of magnitude larger than $k_{T(S)}$, validating the assumption Eq. (6).

In the present study, since [E] is typically two orders of magnitude larger than [S] *and* since the triplet lifetime of the emitter (**2**) in polar media (~ 5 ms[22]) is an order of magnitude larger than that of the sensitizer (**1**) (~ 260 μs[21]), this assumption is to assume that $[^3E^*] >> [^3S^*]$. Based on the same discussion, it is also derived that $k_{T(E)}\,[^3E^*]$ is still a few orders of magnitude larger than $k_{T(S)}\,[^3S^*]$ despite $k_{T(S)} > k_{T(E)}$. Hence, the order-of-magnitude estimation reduces Eq. (3) to

$$N_{ex} = k_{T(E)}[^3E^*] + 2k_{TTA}[^3E^*]^2. \tag{8}$$

Eq. (8) is readily solved to be

$$\sqrt{2k_{TTA}}\,[^3E^*] = \frac{1}{2}\left(\sqrt{\frac{k_{T(E)}^{\;2}}{2k_{TTA}} + 4N_{ex}} - \frac{k_{T(E)}}{\sqrt{2k_{TTA}}}\right). \tag{9}$$

In the parentheses of Eq. (9), the negative sign has to be chosen because of the physical requisite, $[^3E^*] \to 0$ as $N_{ex} \to 0$. By substituting Eq. (9) into Eq. (5), the UC-QY is derived as

$$\frac{\theta}{\varepsilon\varphi} = 1 + \frac{k_{T(E)}^{\;2}}{4k_{TTA}N_{ex}}\left(1 - \sqrt{1 + \frac{8k_{TTA}N_{ex}}{k_{T(E)}^{\;2}}}\right). \tag{10}$$

Here, dimensionless variables Θ and Λ are introduced with a new constant $\alpha$ (in unit of $M^{-1}$ s) as follows

$$\Theta \equiv \frac{\theta}{\varepsilon\varphi}, \quad \Lambda \equiv \alpha N_{ex}, \quad \alpha \equiv \frac{4k_{TTA}}{k_{T(E)}^{\;2}}. \tag{11}$$

Finally, Eq. (11) is rewritten in terms of these variables as

$$\Theta = 1 + \frac{1 - \sqrt{1 + 2\Lambda}}{\Lambda} \quad \left(\Leftrightarrow \frac{\theta}{\varepsilon\varphi} = 1 + \frac{1 - \sqrt{1 + 2\alpha N_{ex}}}{\alpha N_{ex}}\right). \tag{12}$$

This dimensionless equation describes the relationship between $\theta$ and $N_{ex}$ *in terms of two independent linear scaling factors*, $\varepsilon\varphi$ (scaling along the vertical axis) and $\alpha$ (scaling along the horizontal axis). By applying L'Hopital's rule, Eq. (12) has been confirmed to satisfy the physical requisites, "$\Theta \to 0$ as $\Lambda \to 0$" and "$\Theta \to 1$ as $\Lambda \to \infty$".

**Figure 5b** shows the fitted curves from Eq. (12) to the experimental data of $N_{ex}$ vs. $\theta$. The values of $N_{ex}$ were calculated based on the laser beam spot (0.8 mm), the optical path length (1 mm), the molar density of the sensitizer, and the sample's absorbance at the excitation wavelength measured with a 1mm-thick quartz cuvette.[23] The curves predicted by the model (Eq. (12)) show good agreements with the experimental data, corroborating the validity of the assumption (Eq. (6)) in this study.

The scaling factors $\varepsilon\varphi$ and $\alpha$ that provide the best fitting are summarized in **Table. 4**. The magnitude of $\varepsilon\varphi$ is remarkably dependent on the IL, which determines the upper limit of the UC-QY for each case. Assuming the reported value of $\varepsilon \sim 0.87$ for the case of perylene (**2**) in polar solvents,[22] the value of $\varphi$, which is the branching ratio for the singlet formation ($^3E^* + {}^3E^* \rightarrow {}^1E^*$) in TTA process, may have been as high as 0.129/0.87 = 14.8 % for the sample #7. It is noted that Chen et al.[16] have recently reported the experimentally determined values of $\varphi$ to be as high as 33 %, by exciting their samples with intense femto-second laser pulses. This means that the value of $\varphi$ in TTA-UC processes could be higher than 1/9, which is the value derived when the process is purely dictated by the quantum-statistical partitioning condition (i.e., when the formation of quintet states are fully allowed and also diffusion controlled). When the TTA outcomes are limited to singlet and triplet, the value of $\varphi$ can be as high as 40 % in case their branching ratio is assumed to be 1:3.[24] Therefore, the above discussion that $\varphi$ in the present study may have been as high as 14.8 % does not violate previous findings. For the full elucidations of the underlying mechanisms that dictate the magnitude of $\varepsilon\varphi$, detailed studies on the photochemical dynamics of $\varepsilon$ and $\varphi$, perhaps supported by quantum-chemical computational investigations, are needed and currently under investigation.

Finally for this section, it is mentioned that the order-of-magnitude estimation performed above for Eq. (6) is *conservative*, since the actually measured values of diffusion controlled rate constants ($k_{dif}$) in many ILs (including those used in this study) are known to be about an order of magnitude *higher* than the values estimated using Eq. (7).[25] This also supports the above discussion that ILs are not as viscous of a media as they are generally believed. From the above analysis and discussion, it has been shown that *ionic liquids, often regarded as viscous media, are not viscous media for the purpose of TTA-UC*, based on that the condition for an efficient donor-accepter energy transfer (Eq. (6)) is satisfied in this study, corroborated by the agreement of the experimental data and the model derived based on this assumption.

## 4. Conclusions

Phase-stable organic photon upconverters have been fabricated using ionic liquids as the fluidic media for triplet energy transfer processes (Figure 2). It has been found that the polycyclic aromatic molecules used for the triplet sensitizing and triplet-triplet annihilation are stably solvated in the ionic liquids for a long time (Figure 3). The cation-π interaction has been proposed as the mechanism for the stable solvation of the polycyclic aromatic molecules. The upconversion quantum yields have been measured for several ionic liquids (Table 2). It has been found that the quantum yields do not correlate with the viscosity and are remarkably dependent on the ionic liquid used. An upconversion quantum yield as high as 10 % has been achieved, which is much higher than the previous values reported for continuous wave excitations of the samples fabricated with volatile solvents.

The dependence of the upconversion emission intensities on the excitation power was measured, and it revealed that the upconversion quantum yield saturates to a certain limit depending on the ionic liquid (Figure 4). This saturation behavior has been explained in terms of efficient energy transfers between the triplet states. The direct reason for the efficient energy transfers has been attributed to the suppressed oxygen concentrations in the samples, which was made possible with the thorough pumping of the samples with an ultra-high vacuum turbo-molecular pump exploiting the negligible vapor pressures of ionic liquids. From the analysis, it has been shown that ionic liquids are actually not viscous media for triplet-triplet annihilation based photon upconverters. The model (Equation 12) showed good agreement with the experimental results (Figure 5b), corroborating the above discussions.

In essence, ionic liquids have been found to be useful and are proposed as a fluidic media for triplet-triplet annihilation based photon upconverters. The core advantage of using ionic liquids lies in the non-volatility and non-flammability, important properties for practical applications. As the number of ionic liquids known thus far exceeds 1,000,000,[5b] further exploration for optimal ionic liquids is expected to give rise to further enhancement of the upconversion quantum yield.

## 5. Experimental

*Measurement of the UC-QY:* The UC-QYs were measured using the following procedure. For this purpose the ILs #1 – #7 provided from IoLiTec were chiefly used, while those provided from Covalent Associates were used mainly for the purpose of occasional checking of the result. Square cross-sectional quartz glass tubes were purchased from VitroCom (QS101, outer and inner dimensions are 2 × 2 mm and 1 × 1 mm respectively) and cut into 25 mm length and washed. Their one end was closed with a burner. After the sample liquids were evacuated by a turbo molecular pump (Pffeifer, HiCube80) for at least 12 h in a high-vacuum chamber installed inside of a SUS vacuum glovebox (UNICO, UN-650F), the chamber was opened in the glovebox under Ar, as described in Section 2.1. The sample liquids were then injected into the quartz glass tubes to approximately 3/4 of the length using a purpose-made SUS syringe needle, and the tubes' open ends were firmly sealed with Zn/Pb alloy low melting point solder, all of which were done under the Ar-filled SUS glovebox. Similarly, reference liquid, 9,10-bis(phenylethynyl)anthracene (BPEA) dissolved in toluene ($10^{-5}$ M), was sealed in the same quartz tube.

The liquid-containing quartz tube was rigidly held by a custom-designed SUS mounting block installed on a micrometer-actuated SUS precision XYZ stage, in order to assure the accuracy and reproducibility of the spatial positioning of the tube. The sample and the reference tubes were irradiated with a 632.8 nm CW HeNe laser (Melles Griot, 25-LHP-928) and a 407 nm CW diode laser (World Star Tech, TECBL-30GC-405), respectively. The beam spot diameters at the sample were found to be ~ 0.8 mm using a CCD laser beam profiler (Ophir, SP620). The beam line paths for both 632.8 and 407 nm lasers were carefully aligned and made to be identical using two mechanical irises installed in the optical path. The photoemission was collected in the orthogonal direction to the incident laser beam direction. The emission was collimated by a $CaF_2$ lens and then re-focused by a BK7 glass lens onto the entrance slit of a 30 cm monochromator (PI Acton, SP2300). The spectrum was recorded by a thermoelectrically cooled 1340 × 100 pixel Si CCD detector (Princeton Instruments, PIXIS:100BR). The acquired spectrum data was corrected for the wavelength dependences of both the grating efficiency and the CCD detector's sensitivity. Along with this, optical absorption spectra of both the sample and the reference, held in a 1 mm thick quartz cuvette (Starna, Type 53/Q/1), were measured using a UV-vis spectrophotometer (Shimadzu, UV-3600).

The values of UC-QY were calculated based on a standard formula with the known emission quantum yield of BPEA (85 %) [26], the same method as that employed in the previous UC-TTA study [3c]. It is noted that the data analysis in this study has been facilitated by the coincidences in the container material (both quartz) and in the optical path lengths of the photoemission and absorption measurements (both 1 mm). Definition of the UC-QY in this study is same as that in the previous studies; namely, 100 % UC-QY denotes the situation that all of the absorbed photons are upconverted to generate the half number of the absorbed photons.

**Acknowledgements**

The author thanks Profs. Isao Sato, Osamu Ishitani, Akio Kawai, and Tomokazu Iyoda at Tokyo Inst. Tech. for their valuable comments. This work was financially supported by the MEXT-JST program "Promotion of Environmental Improvement for Independence of Young Researchers". The findings in this paper have been documented as a patent 2010-230938JP (pending). Supporting Information is available online from Wiley InterScience or from the author.

Received: ((will be filled in by the editorial staff))
Revised: ((will be filled in by the editorial staff))
Published online: ((will be filled in by the editorial staff))

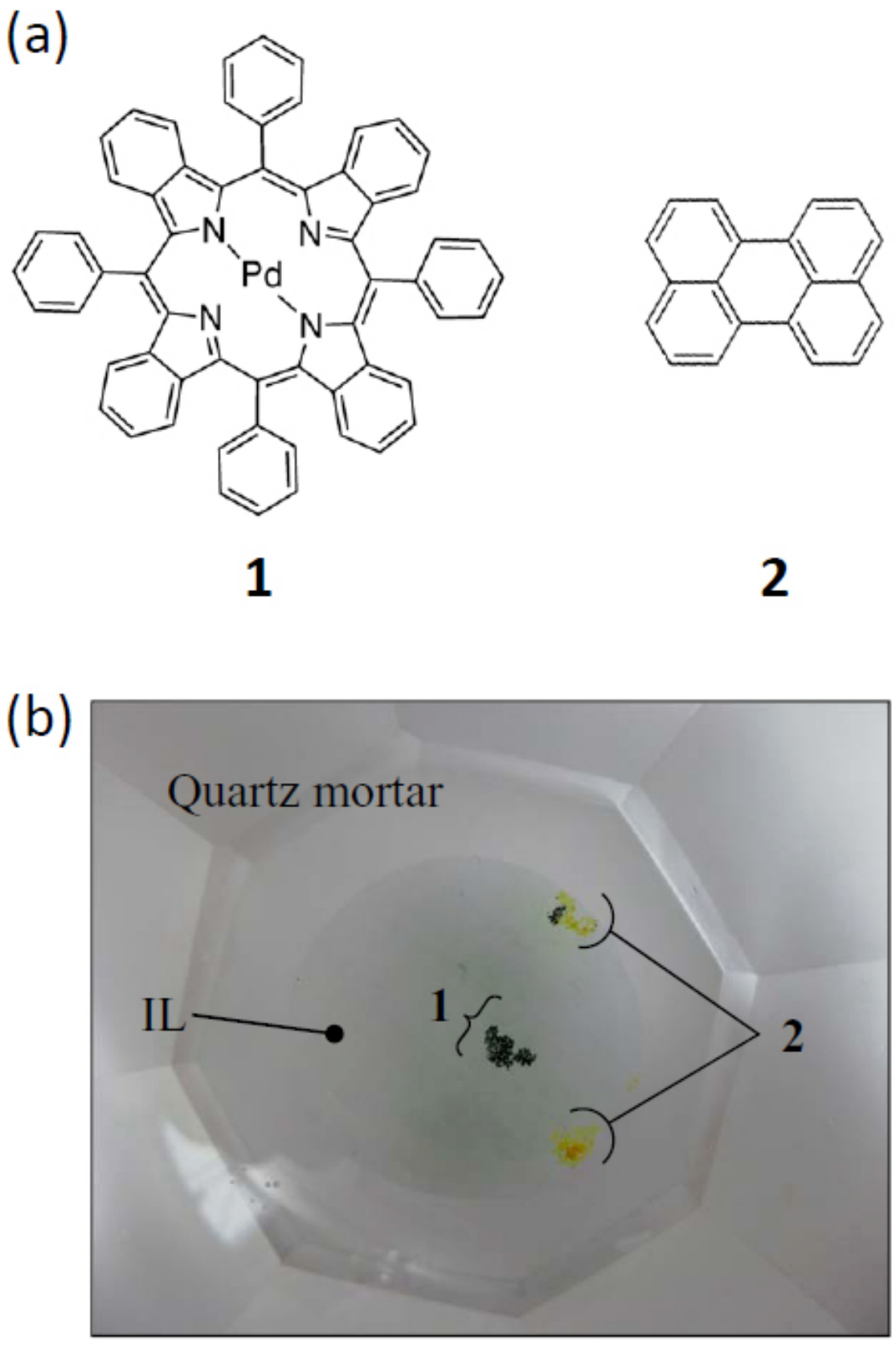

**Figure 1.** (a) The sensitizer ($PdPh_4TBP$, **1**) and the emitter (perylene, **2**) used in this study. (b) A photograph taken 24 h after **1** and **2** were sprinkled over ionic liquid (IL #1) held in the bottom of a quartz mortar.

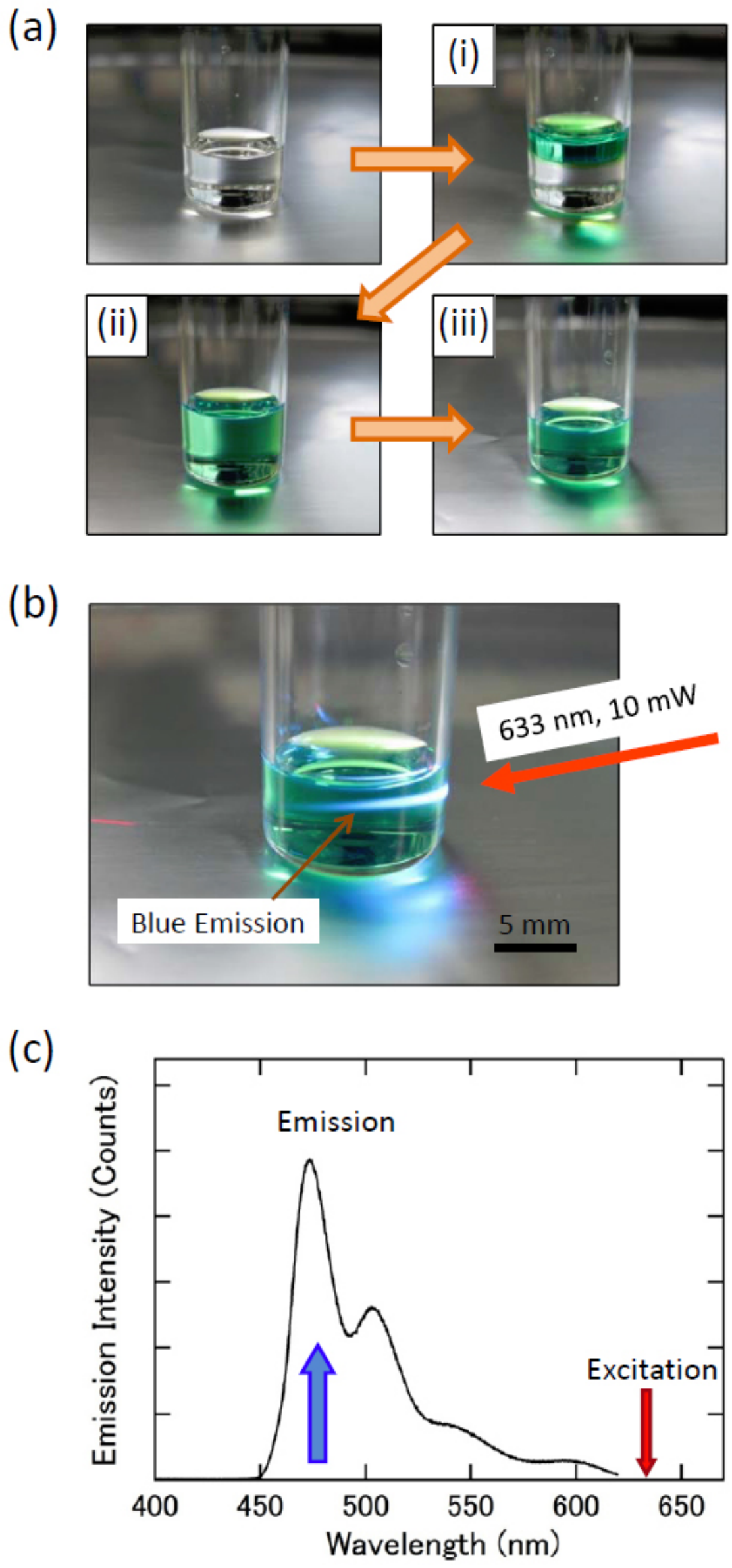

**Figure 2.** Sample made with IL #1 with sensitizer and emitter concentrations of $5 \times 10^{-5}$ M and $1 \times 10^{-3}$ M, respectively. (a) Photographs taken at each step of the sample fabrication. (b) A typical looking photograph of an upconversion of 632.8 nm incident CW light (10 mW) under room light illuminations. (c) Photoemission spectrum.

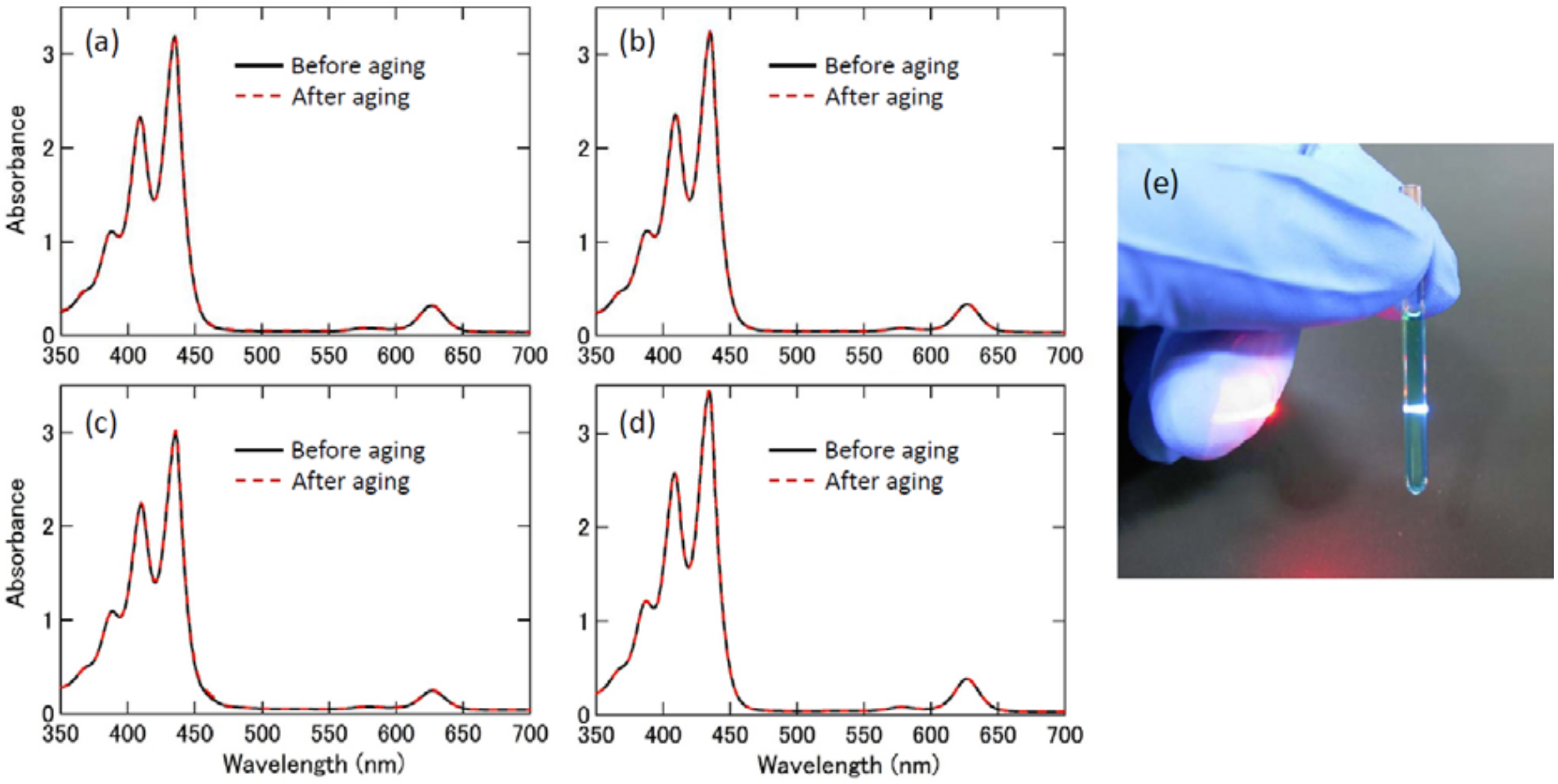

**Figure 3.** (a – d) Comparisons of the optical absorption spectra taken before and after the aging process (see the main text) for the samples fabricated with (a) IL #1, (b) IL #2, (c) IL #6, and (d) IL #8. The sensitizer and the emitter concentrations were $5 \times 10^{-5}$ M and $1 \times 10^{-3}$ M, respectively. Absorption features in the 350 – 450 nm and 550 – 650 nm ranges are of the emitter and the sensitizer, respectively. The differences in the absorbance magnitude among these spectra were caused by the difficulty in accurately pipetting small volumes of the stock solutions. (e) A photo of an upconversion of 632.8 nm CW laser light (10 mW) taken 6 months after its fabrication. The sample had been firmly sealed with low melting point solder in a square cross-section quartz tube (inner: $2 \times 2$ mm, outer: $3 \times 3$ mm, length: 40 mm) .

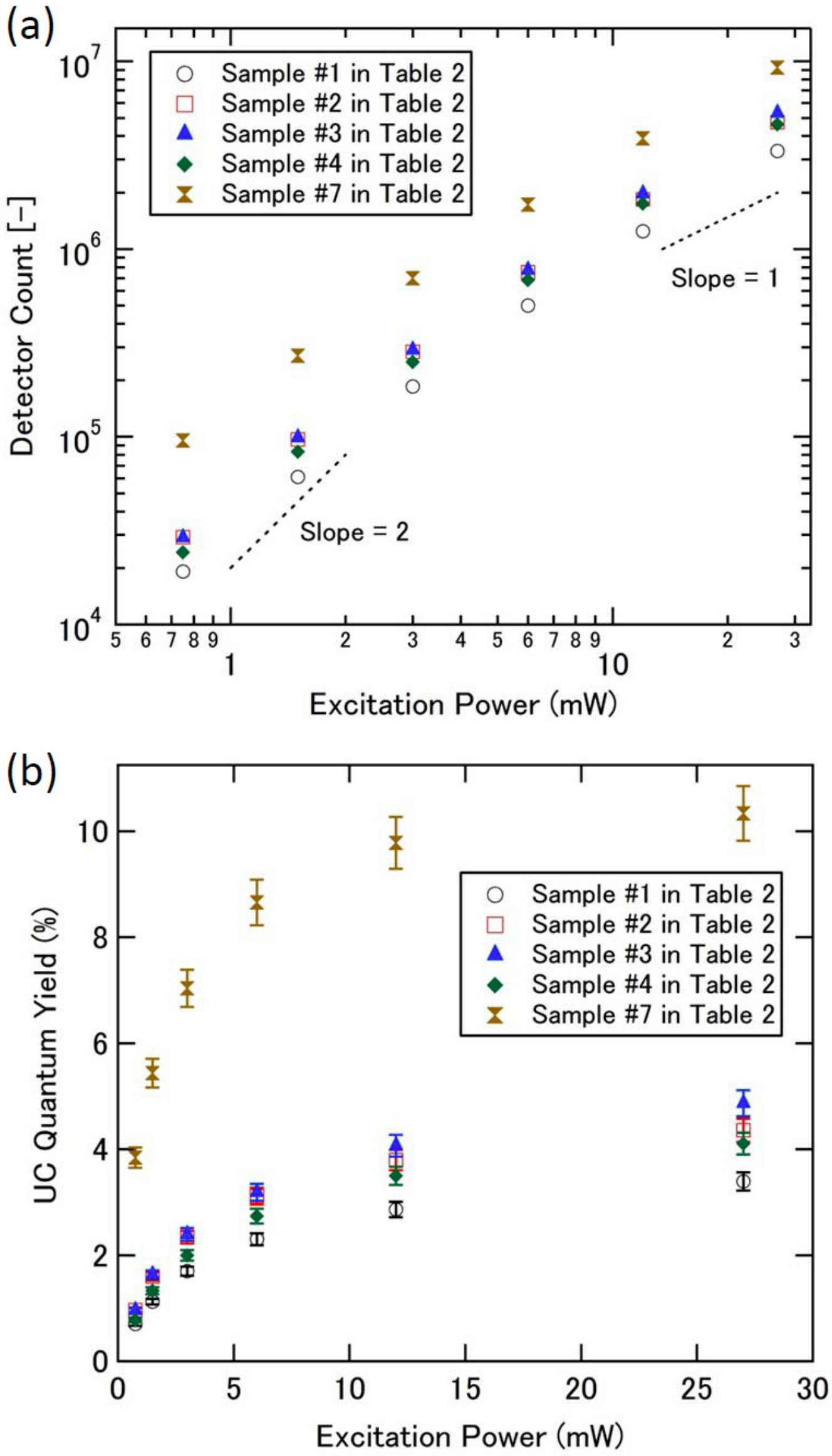

**Figure 4.** (a) Dependence of the upconversion emission intensities on the excitation power of a 632.8 nm CW laser light for the samples fabricated with ILs #1 – #4, and #7, plotted in double logarithmic scale. Dashed lines represent linear and quadric increments shown for eye-guides. (b) Dependence of the UC-QY on the excitation power of the 632.8 nm CW laser light for the same samples. The error bars account for the ±5% certainty in the photoemission measurement.

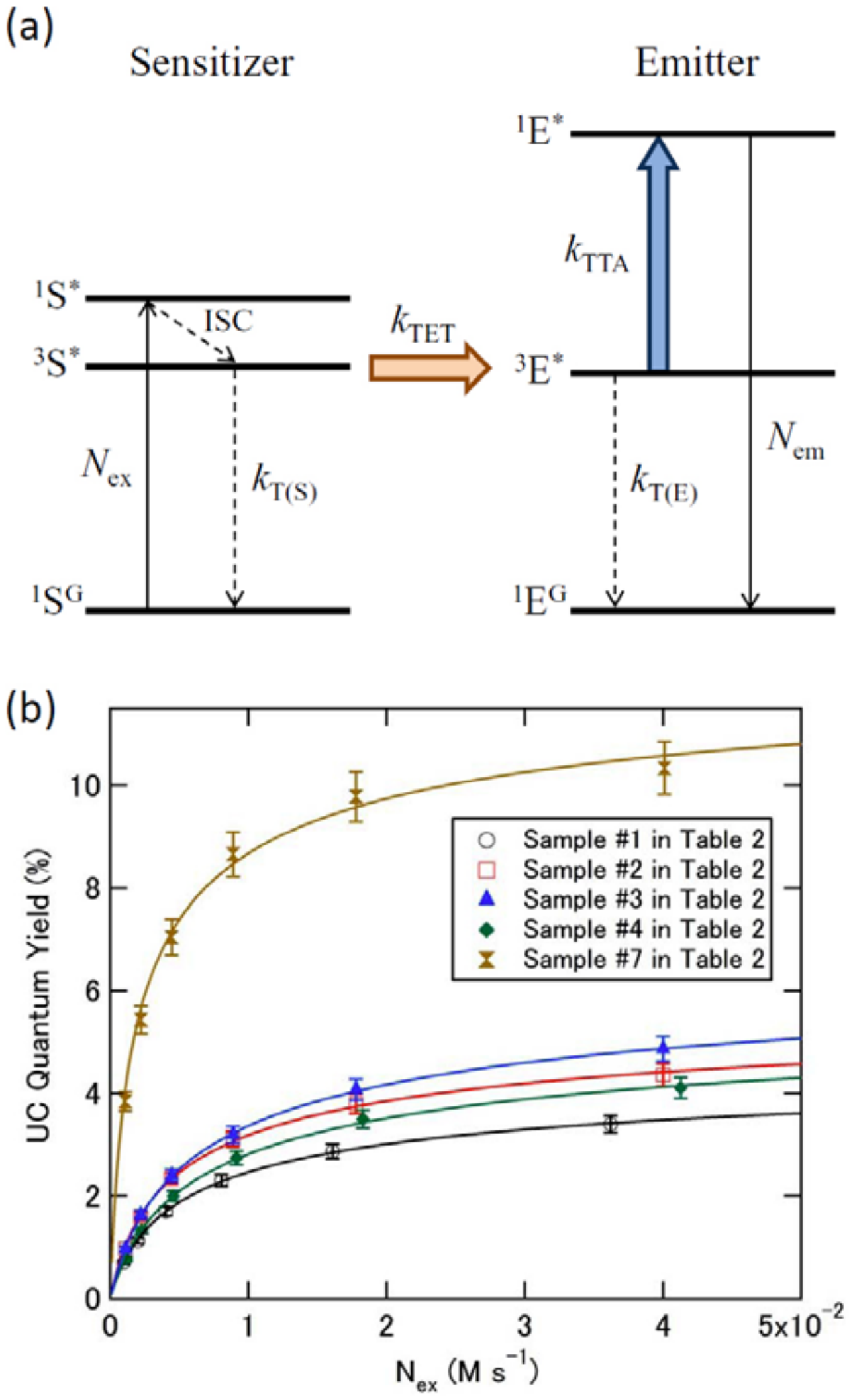

**Figure 5.** (a) Schematic of the energy level diagram in the present system. See the main text for the definitions of the symbols. (b) Plot of the experimental UC-QY vs. $N_{ex}$ relations for the same data set as those shown in Figure 4(b) and the curves fitted to them by Eq. (12). The error bars account for the ± 5% measurement certainty.

**Table 1.** List of the ionic liquids tested.

| # | Ionic Liquid | Mixture Uniformity |
|---|---|---|
| 1 | $[C_2mim][NTf_2]$ | Yes |
| 2 | $[C_4mim][NTf_2]$ | Yes |
| 3 | $[C_6mim][NTf_2]$ | Yes |
| 4 | $[C_8mim][NTf_2]$ | Yes |
| 5 | $[C_2dmim][NTf_2]$ | Yes |
| 6 | $[C_3dmim][NTf_2]$ | Yes |
| 7 | $[C_4dmim][NTf_2]$ | Yes |
| 8 | $[NC_1C_2C_2(C_2OCH_3)][NTf_2]$ | Yes |
| 9 | $[C_3dmim][CTf_3]$ | Yes |
| 10 | $[C_2mim][CH_3CO_2]$ | No |
| 11 | $[C_2mim][CF_3CO_2]$ | No |
| 12 | $[C_3mim][I]$ | No |
| 13 | $[NC_1C_2C_2(C_2OCH_3)][BF_4]$ | No |

**Table 2.** Dependence of the UC-QY on the ionic liquids measured by 632.8 nm CW excitations (30 mW) for the samples fabricated with [S] = $1 \times 10^{-5}$ M and [E] = $3 \times 10^{-3}$ M.

| # | Ionic Liquid | QY(%) |
|---|---|---|
| 1 | $[C_2mim][NTf_2]$ | 3.3 |
| 2 | $[C_4mim][NTf_2]$ | 4.4 |
| 3 | $[C_6mim][NTf_2]$ | 5.2 |
| 4 | $[C_8mim][NTf_2]$ | 4.2 |
| 5 | $[C_2dmim][NTf_2]$ | ~ 1 |
| 6 | $[C_3dmim][NTf_2]$ | ~ 1 |
| 7 | $[C_4dmim][NTf_2]$ | 10.6 |
| 8 | $[NC_1C_2C_2(C_2OCH_3)][NTf_2]$ | 2.5 |

**Table 3.** Dependence of the UC-QY on the sensitizer and emitter concentrations measured by 632.8 nm CW excitations (27 mW) for the samples fabricated with IL #3.

| Sensitizer (M) | Emitter (M) | QY (%) |
|---|---|---|
| $3 \times 10^{-5}$ | $5 \times 10^{-4}$ | 1.9 |
| $1 \times 10^{-5}$ | $1 \times 10^{-3}$ | 4.3 |
| $3 \times 10^{-5}$ | $1 \times 10^{-3}$ | 3.1 |
| $1 \times 10^{-4}$ | $1 \times 10^{-3}$ | 1.4 |
| $3 \times 10^{-5}$ | $3 \times 10^{-3}$ | 4.4 |

**Table 4.** Scaling factors $\varepsilon\varphi$ and $\alpha$ that provided the best fittings by Eq. (12) in Figure 5b.

| # | $\varepsilon\varphi$ ($\times 10^{-2}$) | $\alpha$ ($\times 10^{2}$ $M^{-1}$ s) |
|---|---|---|
| 1 | 5.0 | 3.9 |
| 2 | 6.2 | 4.4 |
| 3 | 7.1 | 3.4 |
| 4 | 6.1 | 3.1 |
| 7 | 12.9 | 12.4 |

**Supporting Information:**

Miscibility of Benzene and Cyclohexane with the Ionic Liquids

Benzene and cyclohexane were purchased from Wako Chemicals. 300 μl of IL #1, IL #3, IL#7, and IL #8 (see Table 1 in the main text) were held in two sets ($2 \times 4 = 8$) of glass crimp vials, and then sufficient amounts (~ 1 ml) of benzene or cyclohexane were added to each set. The vials were capped with silicone rubber septum using a hand crimper and underwent hand-shaking to mix. The resultant look of the vials are shown in Figure 1S. All of them are separated in two distinct layers, which correspond to the IL that have absorbed and saturated with the organic solvents (bottom layer) and the excess amounts of the organic solvents (upper layer). In the figure, the original surface levels of the IL before adding benzene and cyclohexane are indicated by dotted arrows, and the positions of the interfaces after their addition and mixing are indicated by solid arrows. While benzene is miscible with those ILs by finite amount, cyclohexane is completely immiscible with the ILs. While both benzene and cyclohexane are common non-polar solvents, there exists a difference in the miscibility with the ILs.

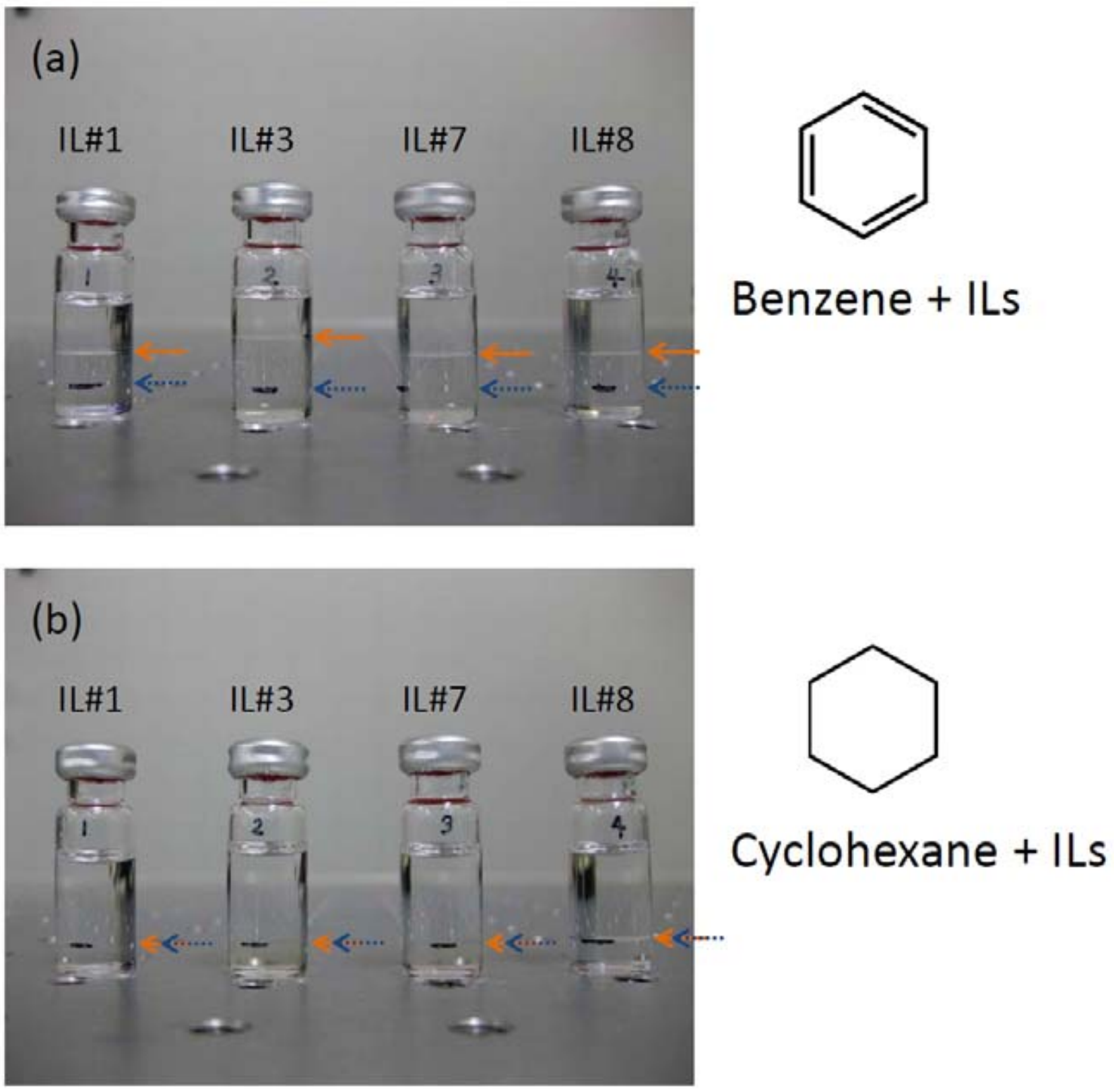

**Figure S1.** Photos of the sample vials taken after mixing of the ILs with (a) benzene and (b) cyclohexane. The dotted arrows indicate the original levels of the ILs in the vials, and the solid arrows indicate the positions of the interfaces between the two layers after mixing.